# Real-time polarization tuning in Mach-Zehnder interferometer using electro-optically modulated twist angles of nematic liquid crystal

Rajneesh Joshi [1, †], Gyaprasad [2,†,*]

[1] Department of Physics, Government Degree College Danya, 263622 Almora, Uttarakhand, India
[2] Department of Physics, Government Degree College Jaithra, 207249 Etah, Uttar Pradesh, India
[†] Both authors contributed equally.
[*] **Corresponding author:** gyaprasadamu@gmail.com

**Abstract:**

We propose a theoretical framework to dynamically control the degree of polarization of light by using the superposition of incoherent orthogonally polarized beams in a Mach-Zehnder interferometer incorporating a twisted nematic liquid crystal cell in one of its arms. The liquid crystal acts as an elecro-optically controlled polarization rotator, where the applied electric field changes the twist of molecules inside the nematic liquid crystal, thereby altering the plane of polarization. This controllable voltage dependent polarization rotation causes manipulation of the output degree of polarization. The resulting system allows real-time, tunable control over the degree of polarization, offering advantages over traditional static or reflection-based approaches, which often suffer from intensity losses or manual errors. We also observe that in the interference of fully coherent orthogonally polarized beams through a similar configuration, the degree of polarization is always equal to 1, whereas the orientation of linear state of polarization is changed with voltage.

## 1 Introduction

Liquid crystals (LCs) represent a unique phase of matter that exists between conventional solid crystals and isotropic liquids [1, 2]. This intermediate state endows them with distinctive optical and electro-optical properties, making them highly valuable in various technological applications. Among different types of LCs, nematic liquid crystals (NLCs) are particularly significant due to their birefringence, which arises from the anisotropic alignment of rod-like molecules along a preferred direction known as the director. The degree of anisotropy depends on molecular orientation—highly aligned molecules result in pronounced birefringence, whereas a lack of alignment leads to near-isotropic behavior [1–4]. NLCs are widely used in display technology [5], adaptive optics, and photonic devices due to their tunable optical properties. By applying an external voltage, the molecular orientation of NLCs can be precisely controlled, leading to changes in their effective refractive indices [6, 7]. This induced realignment generates electric dipoles and modifies the polarization state of incident light, making LCs a cost-effective and versatile alternative to bulk optical crystals. Notably, experiments have demonstrated that NLCs often exhibit superior birefringence characteristics compared to traditional crystalline materials [8–10]. The ability to precisely manipulate the polarization state of light is a crucial asset in numerous scientific and technological domains, particularly in scenarios where controlled adjustment of the degree of polarization (DoP) is essential. This capability is highly advantageous for applications that demand either partially polarized light or a specifically tailored polarization state [11–13]. Tunable control over DoP is integral to various fields, including statistical optics, remote sensing, and polarization imaging systems such as polarization- sensitive cameras [14].

Moreover, dynamic control of DoP has emerged as a vital tool in biomedical optics [15], playing a critical role in non-invasive diagnostics, high-contrast imaging, and tissue characterization. In medical applications, the ability to modify polarization states enhances image clarity, improves contrast in biological imaging, and facilitates early detection of pathological conditions. This is particularly useful in optical coherence tomography

[16], low-coherence interferometry, and laser-based diagnostic techniques [17]. Beyond biomedical fields, tunable polarization sources [18–23] are also instrumental in atmospheric science, where they help analyze aerosols, cloud properties, and environmental pollution levels [24, 25]. In quantum optics, precise polarization control is fundamental to quantum key distribution and secure communication [26, 27]. Additionally, applications in material science, such as stress analysis in transparent materials and defect detection in industrial inspection, greatly benefit from controlled polarization [28].

Over the years, several strategies have been proposed for manipulating DoP, with the majority relying on manually controlled mechanisms or traditional methods which require physical adjustment and are limited in terms of precision, scalability, and responsiveness. Use of electro-optically controlled LCs devices enable real-time, dynamic modulation of the DoP through electronic signals, offering superior advantages in terms of speed, precision, automation potential, energy efficiency, and long-term operational stability. Efforts have done to explore electro-optically driven DoP control using LCs, those methods often depend on the reflection of light [29], depolarization based [30] which inherently suffers from reduced intensity or decoherency, thereby affecting the quality and consistency of polarization manipulation.

In this study, we present a theoretical framework for tuning the DoP of light using the superposition of incoherent orthogonal light fields in a Mach–Zehnder Interferometer (MZI) integrated with a twisted nematic liquid crystal (TNLC) cell in one of its arms. The TNLC acts as a voltage-controlled polarization rotator. When an external voltage is applied across the TNLC cell, the initial helical orientation of the LCs molecules becomes progressively twisted. This voltage-induced realignment alters polarization plane of the incident light, thereby modulating its polarization characteristics. As a result, the system enables precise manipulation of output polarization characteristics and hence the precise tuning of the output DoP by varying the applied voltage across TNLC. This approach provides a robust method for the controllable manipulation of the polarization state of light, which may find applications in quantum communication, optical signal processing, and adaptive polarization-based imaging systems. We also note that, in the superposition of fully coherent orthogonal light fields, the DoP at the output of MZI is always 1, regardless of the voltage applied to the TNLC. However, the orientation of linear polarization changes with voltage across the TNLC.

## 2 Formulation of Polarization Theory

Polarization indicates the direction of the electric field in an electromagnetic wave, which can be determined using important mathematical tools such as the polarization matrix and Stokes parameters. Using these tools, one can determine the correlation between mutually orthogonal polarizations known as the DoP. Generally, the electric field vector at any spatial position ($\boldsymbol{r}$) and time instant $t$ can be expressed in terms of orthogonal polarization components as

$$\boldsymbol{E}(\boldsymbol{r},t) = \begin{bmatrix} E_x(\boldsymbol{r},t) \\ E_y(\boldsymbol{r},t) \end{bmatrix}. \tag{1}$$

All the possible field correlations at the same position and time instant are described by $2 \times 2$ matrix, named the polarization or coherency matrix. It is defined by

$$J(\boldsymbol{r},t) = \begin{bmatrix} J_{xx}(\boldsymbol{r},t) & J_{xy}(\boldsymbol{r},t) \\ J_{yx}(\boldsymbol{r},t) & J_{yy}(\boldsymbol{r},t) \end{bmatrix}, \tag{2}$$

where $J_{ij}(\boldsymbol{r},t) = <E_i^*(\boldsymbol{r},t)E_j(\boldsymbol{r},t)>$, $(i,j = x,y)$ denotes the elements of the polarization matrix. The principal diagonal elements represent the intensities of horizontal and vertical polarizations, whereas the off-diagonal elements represent the $x$–$y$ and $y$–$x$ correlations, which are denoted by $B_{xy}$ and $B_{yx}$, respectively. For convenience, $B_{xy}$ and $B_{yx}$ are written in place of $J_{xy}$ and $J_{yx}$, respectively, for the interfering orthogonal beams in the MZI. Both are complex conjugate to each other, i.e., $B_{yx} = B_{xy}^*$. The polarization states of light are described by the Stokes parameters, which are defined in terms of the polarization matrix as follows:

$$S_0(\boldsymbol{r},t) = J_{xx}(\boldsymbol{r},t) + J_{yy}(\boldsymbol{r},t), \tag{3a}$$

$$S_1(\boldsymbol{r},t) = J_{xx}(\boldsymbol{r},t) - J_{yy}(\boldsymbol{r},t), \tag{3b}$$

$$S_2(\boldsymbol{r},t) = J_{yx}(\boldsymbol{r},t) + J_{xy}(\boldsymbol{r},t), \tag{3c}$$

$$S_3(\boldsymbol{r},t) = i[J_{yx}(\boldsymbol{r},t) - J_{xy}(\boldsymbol{r},t)]. \tag{3d}$$

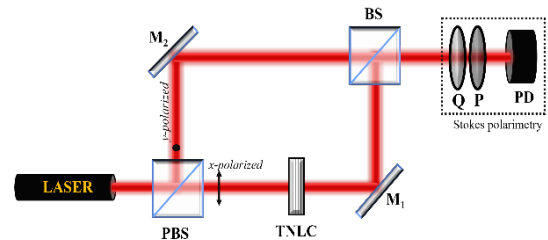

**Fig. 1** Schematic configuration for constructing a tunable source of degree of polarization. PBS polarizing beam splitter, TNLC twisted nematic liquid crystal, $M_1$ and $M_2$ mirrors, BS beam splitter, Q quarter wave-plate, P polarizer, and PD photo diode.

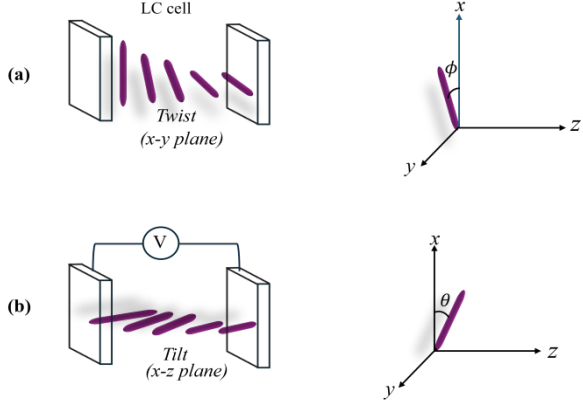

**Fig. 2** Geometry of TNLC cell. (a) shows the twist of molecules inside the LC cell where $\phi$ is twist angle in $x - y$ plane. (b) shows the tilt of molecules in the $x - z$ plane where $\theta$ is tilt angle.

Generally, these correlations are difficult to determine experimentally, and one can also define the Stokes parameters more easily in terms of intensities of different polarization components. These are expressed as follows:

$$S_0(\mathbf{r}, t) = I_x + I_y, \quad (4a)$$

$$S_1(\mathbf{r}, t) = I_x - I_y, \quad (4b)$$

$$S_2(\mathbf{r}, t) = I_{45°} - I_{135°}, \quad (4c)$$

$$S_3(\mathbf{r}, t) = I_{rcp} - I_{lcp}. \quad (4d)$$

The parameter $S_0$ represents the total intensity of the light beam. The $S_1$, $S_2$, and $S_3$ parameters denote the differences in intensities of horizontal and vertical polarization, 45° and 135° polarization, and right and left circular polarization, respectively. The DoP ($P$) quantifies the extent to which light exists in a specific state, such as fully polarized, partially polarized, or unpolarized. For these states, its values are 1, $0 < P < 1$, and 0, respectively. It is a ratio of polarized intensity to the total intensity of a light beam. Using this definition, one can obtain the expression of the DoP in terms of the polarization matrix and Stokes parameters. It is

$$P(\mathbf{r}, t) = \sqrt{1 - 4\frac{\det\{J(\mathbf{r}, t)\}}{[Tr\{J(\mathbf{r}, t)\}]^2}}$$

$$= \frac{\sqrt{S_1^2(\mathbf{r}, t) + S_2^2(\mathbf{r}, t) + S_3^2(\mathbf{r}, t)}}{S_0(\mathbf{r}, t)}. \quad (5)$$

The DoP is directly related to the intensity-normalized orthogonal correlation coefficient, which is defined as

$$\gamma_{xy} = \frac{B_{xy}}{\sqrt{I_x I_y}}. \quad (6)$$

The DoP can be measured experimentally with the help of Stokes polarimetry, as shown in the dotted box in Fig. 1 [31].

## 3. Theoretical Description

The TNLC cell consists of a layer of molecules sandwiched between two substrates with planar alignment. The alignment directions at the two substrates are orthogonal, typically producing a 90° twist in the LCs director from one glass plate surface to the other. The director field at zero applied voltage is described by [1, 3]:

$$n(z) = \begin{bmatrix} \cos\phi(z) \\ \sin\phi(z) \\ 0 \end{bmatrix}, \text{with } \phi(z) = \frac{\pi z}{2d}, 0 \leq z \leq d, \quad (7)$$

where $n(z)$ is the director at position $z$ (along the propagation direction inside the cell), and $d$ is the cell thickness. In this zero-field condition, the director twists uniformly from 0 to $\frac{\pi}{2}$ radians (i.e., 90°). When an external voltage $V$ is applied across the cell (along the z-axis), an electric field $E = \frac{V}{d}$ is established. For LCs with positive dielectric anisotropy $\Delta\varepsilon = \varepsilon_\parallel - \varepsilon_\perp > 0$, the molecules experience a torque that tends to align them with the field, reducing the in-plane twist and causing them to tilt toward the field direction. The elastic and dielectric energies govern the equilibrium configuration of the director field. Under the one-constant approximation of Frank–Oseen theory, the free energy per unit area of the LCs layer is given by [32, 33]:

$$F = \int_0^d \left[ \frac{1}{2} K \left\{ \left(\frac{d\theta}{dz}\right)^2 + \cos^2\theta \left(\frac{d\phi}{dz}\right)^2 \right\} - \frac{1}{2} \varepsilon_0 \Delta\varepsilon E^2 \sin^2\theta \right] dz, \quad (8)$$

where $\theta(z)$ is the tilt angle (angle between the director and the substrate plane, angle in $x$-$z$ plane) and $\phi(z)$ is the twist angle (angle in the substrate plane, angle in $x$-$y$ plane) as illustrated in Fig. 2. $K$ is the Frank elastic constant and $\varepsilon_0$ is the equilibrium configuration is found by minimizing $F$, resulting in the Euler–Lagrange equations:

$$K\frac{d^2\theta}{dz^2} - K\sin\theta\cos\theta\left(\frac{d\phi}{dz}\right)^2 + \varepsilon_0\Delta\varepsilon E^2 \sin\theta\cos\theta = 0, \quad (9)$$

$$\frac{d}{dz}\left(K\cos^2\theta \frac{d\phi}{dz}\right) = 0. \quad (10)$$

These equations are nonlinear and usually solved numerically. However, for practical modeling and optical simulations, an empirical *sigmoidal approximation* is often used to describe the voltage-dependent suppression of the effective twist angle. This sigmoidal model is widely used in modeling TNLC-based optical devices and is supported by foundational works of Gooch and Tarry [34], where full optical and director-based analysis demonstrates a similar S-shaped transition in twist and transmission response with applied voltage. The effective twist angle $\phi_{eff}(V)$, which determines the optical retardation or transmission in devices such as LCDs, can be modeled by a logistic (sigmoidal) function [32,34,35]:

$$\phi_{eff}(V) = \frac{\phi_0}{1+e^{a(V-V_{th})}}, \quad (11)$$

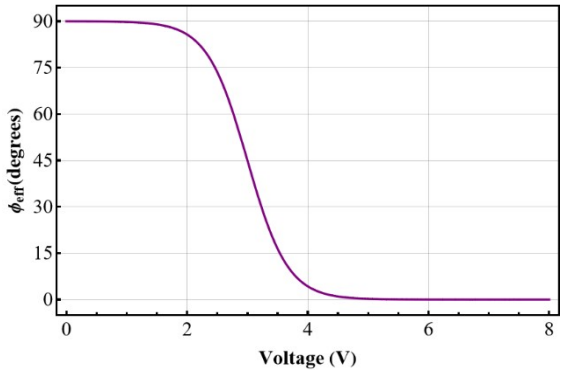

**Fig. 3** The twist angle as a function of applied voltage. As the voltage increases, the twist angle decreases.

where $a$ is the steepness parameter and $V_{th}$ is threshold voltage for the LC cell. This variation of twist angle of LCs with voltage as shown in Fig. 3. The threshold voltage for an LC cell is assumed to be 2 volts; however, it may vary across different cells due to varying laboratory conditions such as temperature, etc. The steepness parameter $a$ is assumed to have a value of 3; however, it may be adjusted based on the experimental data to ensure a more accurate fit to the observed curve. Similarly, the voltage-induced tilt angle can be modeled as:

$$\theta(V) = \theta_{max}\left[1 - \frac{1}{1+e^{a(V-V_{th})}}\right]. \quad (12)$$

These expressions capture the gradual reduction of the twist and increase of the tilt with increasing voltage, starting from the mid-layer and eventually reaching nearly uniform homeotropic alignment at high voltage. At low voltages ($V < V_{th}$), $\theta(z,V) \approx \theta_0\left(1 - \frac{z}{d}\right)$ and at high voltages ($V \gg V_{th}$), $\theta(z,V) \approx 0$, reflecting complete unwinding of twist.

The twist of the LCs molecules governs the rotation of the plane of polarization of the incident light. In the absence of an applied voltage, the initial twist is 90°, which results in a 90° rotation of the polarization plane, as illustrated in Fig. 4. For example, if the incident electric field is initially polarized along the *x*-direction, it becomes *y*-polarized after passing through the cell. When a voltage is applied, the molecular twist gradually decreases, and the rotation of the polarization plane correspondingly reduces. This voltage-dependent control over polarization rotation is utilized in one arm of a MZI to flip the polarization plane of one of the interfering beams. After passing through the arms, the two beams are recombined at a beam splitter. By analyzing the correlation between the orthogonal polarization components at the output, we can calculate DoP as a function of the applied voltage.

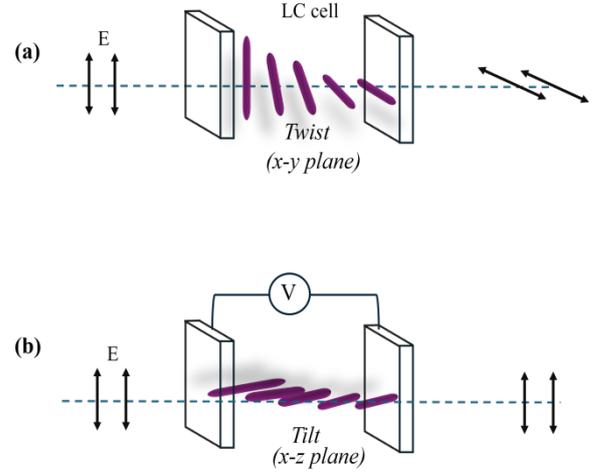

**Fig. 4** (a) The polarization plane of incident electric field is guided by the twist of molecules. (b) At higher applied voltage where twist is completely deformed, polarization plane remains intact.

We now proceed with the following methodology. We use a MZI to control the DoP using a TNLC in one of its arms. Light from a polarized laser is incident on the polarizing beam splitter (PBS); hence, the reflected light represents vertical polarization (*y- polarized*), and transmitted light denotes horizontal polarization (*x- polarized*). A TNLC is placed in the transmitted arm, and both the reflected and transmitted light beams are merged through a nonpolarizing beam splitter. The transmitted and reflected electric field vector can be written as follows.

$$\mathbf{E}_1 = \frac{1}{\sqrt{2}}E_x\hat{x}, \quad (13a)$$

$$\mathbf{E}_2 = \frac{1}{\sqrt{2}}E_y\hat{y}, \quad (13b)$$

where $\hat{x}$ and $\hat{y}$ are unit vectors along the $x$ and $y$ directions, respectively, while $z$ denotes the propagation direction of the light beam. $E_x$ and

$E_y$ represent the horizontal and vertical field components of the electric field vector, which are generally complex numbers. The factor $\frac{1}{\sqrt{2}}$ denotes the transmission and reflection coefficients, respectively, assuming the PBS to be 50 : 50. The plane of polarization of the transmitted component (*x-polarized*) through PBS will be gradually changed after passing through the TNLC cell. The rotation matrix for the light component passing through TNLC is given by

$$J(\phi) = \begin{bmatrix} cos\phi & -sin\phi \\ sin\phi & cos\phi \end{bmatrix}, \quad (14)$$

where $\phi$ denotes the angle by which plane of polarization is rotated which is the twist angle in the TNLC. Now, the transmitted electric field vector through PBS and passes through TNLC cell can be expressed as

$$\boldsymbol{E}_1^t = J[\phi(V)]\boldsymbol{E}_1 = E_x \begin{bmatrix} cos\phi \\ sin\phi \end{bmatrix}. \quad (15)$$

The angle $\phi$ is voltage-dependent and is more accurately represented as $\phi_{eff}(V)$ in Eq. (11). However, for simplicity, we will refer to it as $\phi(V)$ in the subsequent discussion. Therefore, the resultant electric field vector at output of the nonpolarizing beam splitter can be written as

$$\boldsymbol{E}(\boldsymbol{r},t) = \frac{1}{\sqrt{2}}\boldsymbol{E}_1^t + \frac{1}{\sqrt{2}}\boldsymbol{E}_2 = \begin{bmatrix} 1/2\, E_x cos\phi(V) \\ 1/2\, E_x sin\phi(V) + 1/2\, E_y \end{bmatrix}. \quad (16)$$

The $x$ and $y$ components of the resultant electric field vectors in Eq. (16) are

$$E_x(\boldsymbol{r},t) = \tfrac{1}{2} E_x cos\phi(V), \quad (17a)$$

$$E_y(\boldsymbol{r},t) = \tfrac{1}{2} E_x sin\phi(V) + \tfrac{1}{2} E_y. \quad (17b)$$

By employing these field components along Eq. (2), one can calculate the elements of the polarization matrix as

$$J_{xx}(\boldsymbol{r},t) = \tfrac{1}{4} E_x^2 cos^2\phi(V), \quad (18a)$$

$$J_{yy}(\boldsymbol{r},t) = \tfrac{1}{4} E_x^2 sin^2\phi(V) + \tfrac{1}{4} E_y^2 + \tfrac{1}{2} sin\phi(V) Re(B_{xy}), \quad (18b)$$

$$J_{xy}(\boldsymbol{r},t) = \tfrac{1}{4} E_x^2 sin\phi(V) cos\phi(V) + \tfrac{1}{4} B_{xy} cos\phi(V), \quad (18c)$$

$$J_{yx}(\boldsymbol{r},t) = \tfrac{1}{4} E_x^2 sin\phi(V) cos\phi(V) + \tfrac{1}{4} B_{yx} cos\phi(V), \quad (18d)$$

where $B_{xy} = <E_x^* E_y>$ denotes the correlation coefficient between $E_x$ and $E_y$ as previously discussed, $Re$ stands for real part. Hence, the polarization Stokes parameters using Eq. (3a) yield as

$$S_0(\boldsymbol{r},t) = \tfrac{1}{4}(E_x^2 + E_y^2) + \tfrac{1}{2} sin\phi(V) Re(B_{xy}), \quad (19a)$$

$$S_1(\boldsymbol{r},t) = \tfrac{1}{4} E_x^2 cos2\phi(V) - \tfrac{1}{4} E_y^2 - \tfrac{1}{2} sin\phi(V) Re(B_{xy}), \quad (19b)$$

$$S_2(\boldsymbol{r},t) = \tfrac{1}{2} E_x^2 sin\phi(V) cos\phi(V) + \tfrac{1}{2} cos\phi(V) Re(B_{xy}), \quad (19c)$$

$$S_3(\boldsymbol{r},t) = -\tfrac{1}{2} Im(B_{xy}) cos\phi(V), \quad (19d)$$

where $Im$ represents the imaginary part, and $B_{xy} = |B_{xy}| \exp(\iota\delta)$, where $\iota$ represents the imaginary number, and $\delta$ denotes the phase of $B_{xy}$, which is generally assumed to be the phase difference between the arms of the MZI. By applying this property, the DoP using Eq. (5) is obtained as

$$P(\boldsymbol{r},t) = \left[1 - \frac{4\left(I_x I_y - |B_{xy}|^2\right)cos^2\phi(V)}{(I_x + I_y + 2|B_{xy}|sin\phi(V)cos\delta)^2}\right]^{1/2}, \quad (20)$$

where the intensities of horizontal or vertical polarization can be expressed as $I = |E_j|^2$ ($j = x,y$). By substituting the value of $|B_{xy}|$ from Eq. (6) and using the intensity ratio $I_R = \frac{I_y}{I_x}$, we obtain the expression for the DoP as

$$P(\boldsymbol{r},t) = \left[1 - \frac{4 I_R cos^2\phi(V)(1 - |\gamma_{xy}|^2)}{(1 + I_R + 2\sqrt{I_R}|\gamma_{xy}|sin\phi cos\delta)^2}\right]^{1/2}. \quad (21)$$

The above formula represents a general expression for the DoP.

### 4. Results and Discussion

A TNLC is placed in one arm of the MZI, which is the main objective of our study, as it introduces a voltage dependence in the DoP. Equation (21) represents the expression for the DoP at the output of the MZI, which depends on the intensity ratio (or relative intensity) of the interfering orthogonal beams, the twist angle and the intensity-normalized orthogonal correlation coefficient. Now we discuss two following cases:

**(i)** If the two interfering waves with orthogonal polarization are fully coherent, i.e. $|\gamma_{xy}| = 1$, then the orthogonal components do not produce interference fringes; the net intensity at the observation plane is simply the sum of the intensities of the two beams. When a TNLC is placed in the transmitted arm, the linear polarization of that arm is rotated with twist angle $\phi$. As a result, the intensities of the horizontal and vertical components at the output change with the twist angles, while the total intensity remains constant. In this case, by putting $|\gamma_{xy}| = 1$ in Eq. (21), the DoP at the observation plane is calculated as

$$P(r,t) = 1, \tag{22}$$

regardless of the intensity ratio of the orthogonal components and twist angle of TNLC. Although the polarization state itself varies with the intensity ratio and twist angle and remains linear with a changing orientation. The orientation angle of the plane of linear polarization measured with $x$- axis is given by $\psi = \tan^{-1}\sqrt{\frac{I_y}{I_x}}$. Substituting $I_x$ and $I_y$ from Eqs. (17a) and (17b), yields

$$\psi = \tan^{-1}\sqrt{\sec^2\phi(V)(1+I_R) - 1 + 2\sqrt{I_R}\tan\phi(V)\sec\phi}$$

$$\tag{23}$$

Therefore, the orientation angle depends on intensity ratio and twist angle.

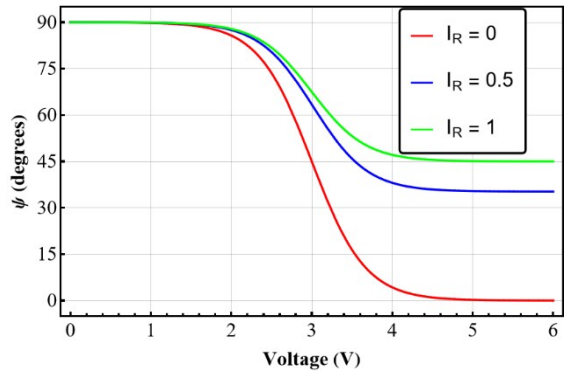

**Fig. 5** Variation of orientation angle of linear polarization with voltage for different intensity ratios $I_R = \frac{I_y}{I_x}$.

For $I_R = 1$, i.e., equal intensities of horizontal and vertical polarization, the orientation angle of linear polarization is given by

$$\psi = \tan^{-1}[\tan\phi(V) + \sec\phi(V)]. \tag{24}$$

Figure 5 shows the variation of $\psi$ (in degrees) as a function of the applied voltage $V$ (in volts). The different curves corresponding to different values of the intensity ratio $I_R$. For $I_R = 0$ (red curve), $\psi$ starts from $90°$ and decreases rapidly to $0°$ as the voltage increases. For $I_R = 0.5$ (blue curve), the transition is smoother, and $\psi$ saturates just above $30°$. For $I_R = 1$ (green curve), the saturation shifts even higher, stabilizing at $45°$.

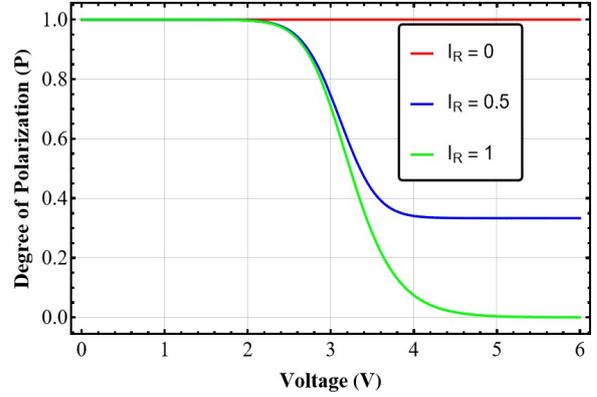

**Fig. 6** Variation of DoP with the applied voltage (in volts) across a TNLC cell for different intensity ratios, $I_R = \frac{I_y}{I_x}$ in the MZI.

(ii) If the interfering orthogonal polarizations are incoherent [path difference is greater than the coherence length of the input polarized source, or path difference does not matter if input light source is unpolarized], i.e., $|\gamma_{xy}| = 0$, then upon interference in the MZI no fringes are formed. In this case, the net intensity is simply the sum of the intensities of the orthogonal polarizations. This net intensity remains constant with respect to the twist angle, although the individual intensities of the orthogonal components vary. We note that the polarization state is not deterministic and becomes fully polarized only at a twist angle of $90°$. Substituting the voltage-dependent phase $\phi(V)$ from Eq. (11), we obtain DoP as

$$P(r,t) = \frac{\sqrt{1+I_R^2 - 2I_R\cos\left[2\phi_0\left(\frac{1}{1+e^{a(V-V_{th})}}\right)\right]}}{1+I_R}.$$

$$\tag{25}$$

We note that, for superposition of incoherent orthogonal polarization, the DoP depends on the intensity ratio of interfering orthogonal polarization and twist angle of LCs.

For equal intensities, i.e., $I_R = 1$, we obtain

$$P(r,t) = \sin\left[\phi_0\left(\frac{1}{1+e^{a(V-V_{th})}}\right)\right]. \tag{26}$$

Figure 6. illustrates the variation of the DoP as a function of the applied voltage across a TNLC cell placed in one arm of the MZI. The plot contains three distinct curves, each corresponding to a different intensity ratio $I_R$. These intensity ratios can be experimentally controlled using various types of PBS. As seen in the Fig. 6, when the intensity ratio is unity ($I_R = 1$), which corresponds to equal intensities in both arms, the DoP exhibits its maximum range, ideally varying from 1 to 0 with applied voltage. This condition represents the most efficient polarization modulation scenario. This is depicted by the green curve in Fig. 6. For a reduced intensity ratio of 0.5 ($I_R = 0.5$), where the intensity in the transmitted arm is twice that in the reflected arm, the DoP starts from 1 and decreases to a value slightly above 0.3 with increasing voltage as depicted by the blue curve. The modulation range is noticeably less compared to the equal-intensity case. When the intensity ratio is zero, meaning the reflected arm is completely blocked and all light propagates through the transmitted arm alone. In this case, no variation in DoP is observed with applied voltage, and it has a maximum value of 1 as depicted by the red curve. This clearly indicates that the presence of non- zero intensity in both arms is essential for any polarization modulation to take place. The most pronounced and desirable modulation behavior is achieved when the intensities in both arms are equal, underlining the importance of balanced optical power distribution for maximizing polarization control.

## 5. Conclusion

In conclusion, we have developed a theoretical framework for controlling the degree of polarization using a Mach-Zehnder interferometer with a twisted nematic liquid crystal. Unlike conventional approaches that rely on mechanical adjustments, static optical elements, or reflection-induced depolarization—which often suffer from reduced intensity or coherence loss—our method utilizes a transmission-based configuration that preserves signal strength while enabling fine control over degree of polarization. The novelty of this study lies in demonstrating an electro-optically controlled, transmission-mode system for degree of polarization modulation using twisted nematic liquid crystal within an interferometric setup. This technique offers several key advantages, including rapid response, energy efficiency, scalability, and compatibility with automated systems. The proposed approach opens new possibilities for advanced optical applications such as adaptive polarization-based imaging, biomedical diagnostics, optical signal processing, and quantum communication. By overcoming limitations of traditional polarization control techniques, our method provides a robust and versatile platform for next generation photonic and quantum technologies.

## 6 Acknowledgment

The authors are grateful to Prof. Bhaskar Kanseri for his invaluable insights, constructive discussions, and conceptual guidance on the optics of liquid crystals in the past.

### Disclosure

The authors declare no conflicts of interest.

### Declaration of competing interest

The authors declare that they have no known competing financial interests or personal relationships that could have appeared to influence the work reported in this paper.